\documentclass[superscriptaddress]{revtex4} 
\usepackage{amsmath,amsfonts,amssymb}
\usepackage{graphicx}
\usepackage{setspace}
\usepackage{multirow}
\usepackage{array, booktabs}
\usepackage{hyperref}
\usepackage{siunitx}

\DeclareSIUnit{\mas}{mas}
\DeclareSIUnit\parsec{pc}
\DeclareSIUnit{\arcsec}{asec}
\DeclareSIUnit{\arcmin}{amin}
\DeclareSIUnit\lightyear{ly}
\DeclareSIUnit{\AU}{AU}

\begin{document} 

\title{Wide Aperture Exoplanet Telescope: a low-cost flat configuration for a 100+ meter ground based telescope}
\author{Benjamin Monreal}
\email{benjamin.monreal@case.edu}

\affiliation{Case Western Reserve University}
\author{Christian Rodriguez}
\affiliation{University of California, Santa Barbara} 

\author{Ama Carney}
\author{Rob Halliday}
\author{Mingyuan Wang}
\affiliation{Case Western Reserve University}

\begin{abstract}
The Wide Aperture Exoplanet Telescope (WAET) is a ground-based optical telescope layout in which {\em one} dimension of a filled aperture can be made very, very large (beyond \SI{100}{\m}) at low cost and complexity.   With an unusual beam path but an otherwise-conventional optics, we obtain a fully-steerable telescope on a low-rise mount with a fixed gravity vector on key components.   Numerous design considerations and scaling laws suggest that WAET can be far less expensive than other giant segmented mirror telescopes.
\end{abstract}

\keywords{Telescope design, observatories, mechanical design, mount}
\maketitle


\begin{spacing}{1}   

There is a strong science case for building ground-based optical telescopes larger than the \SI{30}{\m} ones now under construction.   Some topics, like high-redshift galaxies studies, particularly require larger collecting areas; others, like reflected-light exoplanet detection, benefit from sharp angular resolution even with limited area.   A filled aperture telescope with \SI{100}{\m}-class diffraction limits would give us access to routine reflected-light exoplanet characterization for a large number of systems.   Unfortunately, in the context of observed telescope size/cost scaling laws\cite{stepp_estimating_2003}, which suggest that construction costs increase as $D^{\text{ 2.7--2.5}}$ or $A^{\text{1.35--1.25}}$, \SI{100}{m} circular aperture telescopes are unaffordable.  They require a vast amount of glass (with costs scaling as $A$) and a huge mount and dome (with costs scaling as $A^{\sim 3/2}$) in order to improve the diffraction limit by $\sqrt{A}$.

In this paper, we show a telescope layout with a different scaling law.  We implement a highly elongated pupil, with one long dimension $L$ and one short dimension $W$, which we refer to as WAET (Wide Aperture Exoplanet Telescope).  The total mirror area scales as $\sim$$2.4A$, but the mount/superstructure/dome are extremely simplified, offering costs scaling as $A^{1.0}$ or slower.   We will introduce the general features of the WAET geometry, which may have many different possible implementations.   For concreteness, we will show mechanical and optical design exercises for one implementation (hectometer-WAET or ``hWAET'' at \SI{100 x 2}{\m}) which we argue is a low-risk route towards ground-based direct imaging of rocky exoplanets; and for a much larger instrument (``kWAET'' at \SI{300 x 5}{\m}) with post-TMT-class light collection and sub-milliarcsecond resolution.   

\section{Overview}

In this section, we survey the basic WAET operating principles and preview some of the advantages and disadvantages of the design.  Fig.~\ref{fig1} shows an optical model and Fig.~\ref{fig2} shows some mechanical details to illustrate the beampath.  For concreteness, both figures show Ritchey-Chr\'etien optics but this is not a general requirement.

\subsection{Basic layout} \label{sec_layout}

\begin{figure}
\begin{center}
\includegraphics[width=0.9\textwidth]{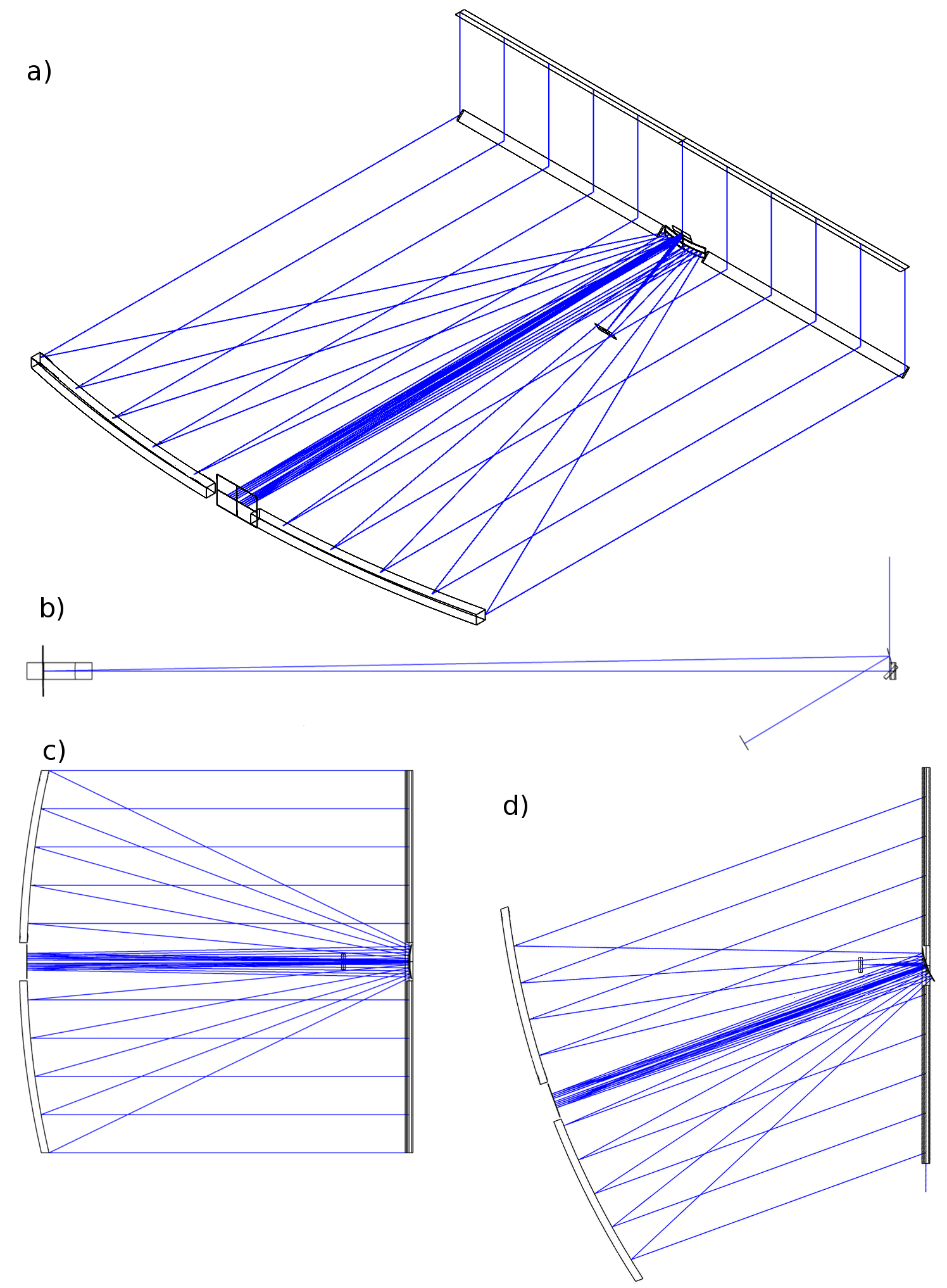}
\end{center}
\caption{WAET optical layout.   Starlight (shown as coming from zenith) reflects once off a tilted siderostat, then off a focusing primary.   The details shown are of an f/1.1 Ritchey-Chr\'etien with a \SI{100 x 2}{\m} rectangular aperture.    a) isometric view. b) side view.   c) top view with the telescope viewing a source at zenith.  d) top view illustrating the ``slew'' positioning of the primary; the telescope is viewing a target 20$^\circ$ north of zenith.}\label{fig1}
\end{figure} 

\begin{figure}
  \includegraphics[width=\textwidth]{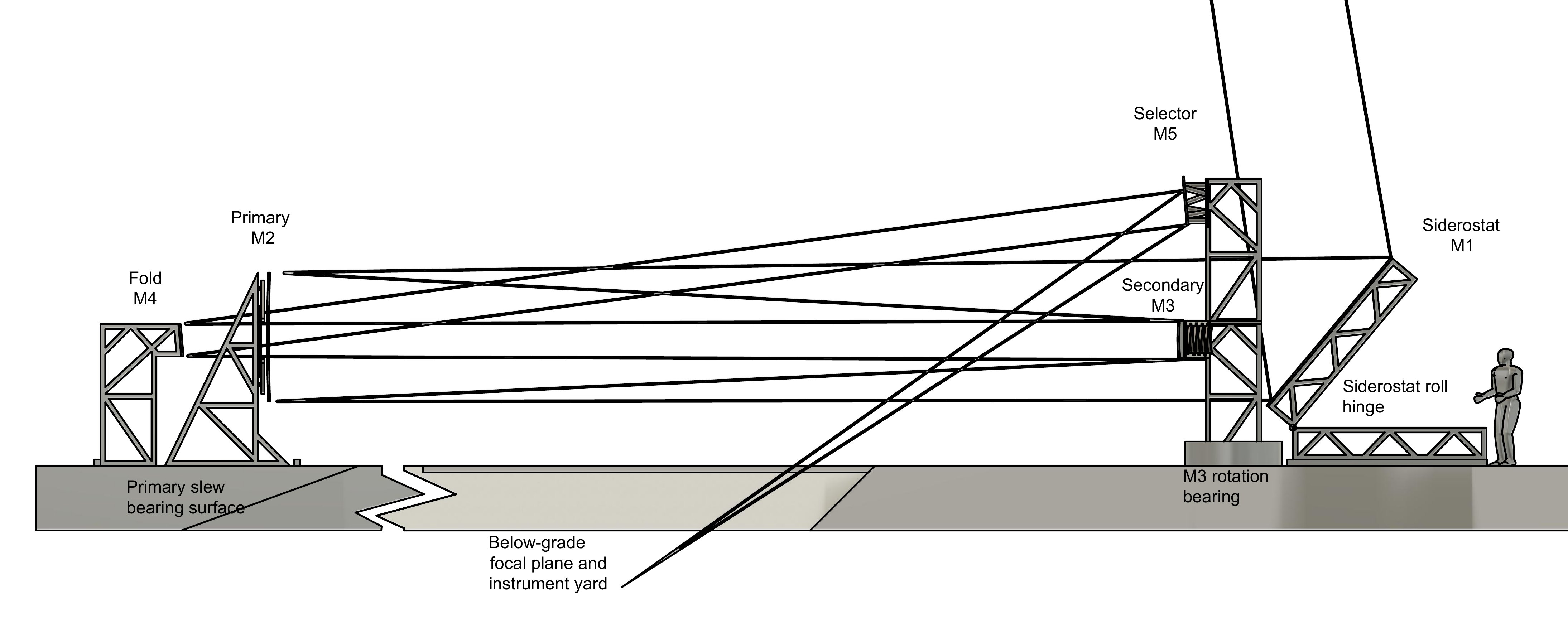}
    \caption{Compressed cross section of WAET, illustrating a folded Richey-Chr\'etien beam path and naming major components.  Components have been repositioned left/right for visibility; for a view of the full installation see Figure \ref{fig_hwaet_mech}.}\label{fig2} 
  \end{figure}  

Starlight is incident on a flat siderostat (M1).   The siderostat is a thin, elongated shape with its longer dimension oriented roughly N-S.  Its only degree of freedom is \emph{roll} around the N-S axis.  The siderostat folds the beam into the horizontal plane and directs it towards the horizon-facing primary (M2).   The primary focusing mirror is, like the siderostat, a thin elongated shape with its short axis vertical and its optical axis parallel to the ground.   The primary mirror moves with one degree of freedom: it can \emph{slew} about a vertical axis near the center of the siderostat.   The prime focus, located at or near this vertical axis, faces horizontally and swivels or rotates with the siderostat slew.   Siderostat-roll and primary-slew, working together, steer the telescope's optical axis freely across most of the sky without the need to elevate any mirrors out of the ground plane.  The two degrees of freedom map to sky coordinates the same way as those of an ``elevation-elevation'' (el-el) telescope mount\cite{manly_unusual_1995}, or for an equatorial mount at an equatoral site.   In a spherical coordinate system $\theta,\phi$ whose symmetry axis is parallel to the ground along the siderostat roll axis, siderostat-roll selects azimuthal angle $\phi$ and primary-slew selects polar angle $\theta$.  

WAET can be seen as a fully-steerable variant of the Kraus-type radio telescope\footnote{Note that our choice of terminology will differ from Kraus, who refers to the siderostat as the ``primary'' and the large focusing element as the ``secondary''.}, historically implemented as the Big Ear at Ohio State (1963--1998)\cite{Kraus:SkyAndTelescope:1953} and the Nan\c{c}ay Radio Telescope \mbox{(1965--)\cite{Lequeux:JournalOfAstronomicalHistoryAndHeritage:2010}}.   In contrast to WAET, Kraus-type telescopes have an east-west long axis and a non-tracking siderostat; they operate at fixed elevation, either as transit telescopes with a moveable (\SI{15}{\degree\per\hour}) feed that can track targets briefly at the chosen elevation.   WAET has its siderostat oriented roughly N-S; targets can be tracked for $\sim$\SI{6}{\hour} by rolling the siderostat quickly (\SI{7.5}{\degree\per\hour}) and slewing the primary slowly ($<$\SI{1}{\degree\per\hour} for most sites and targets).  (Sky coverage will be discussed in detail in section \ref{label_skycoverage}.)   

The system is compatible with various well-understood optical prescriptions (Richey-Chr\'etien, Gregorian, Newtonian, spherical, etc.) so there are no R\&D risks associated with unusual figuring or alignment challenges.   A wide rectangular aperture has certain fabrication advantages, but elliptical or other apertures with more attractive PSFs are also feasible.  In the Richey-Chr\'etien configuration shown in Figs \ref{fig1} and \ref{fig2}, the secondary mirror (M3) slews with the primary, and two flat fold mirrors (M4 fixed in the center of M2, steerable M5 above M3) bring the focal plane to a stationary instrument yard at or below ground level.   WAET is unusual in allowing all instruments, including prime focus instruments, to be stationary and at ground level.

\subsection{Mechanical design and cost} The WAET layout allows us to use extremely simple, lightweight mechanical structures.   The primary mirror is a low-rise, non-tilting, non-wind-loaded structure; it has a constant gravity vector and does not flex (except due to bearing flatness) while tracking.  The siderostat is a low-rise, non-wind-loaded structure; although it does tilt while tracking, first order gravitational flexing, if uncorrected, affects focusing along the low-resolution axis, not the high-resolution axis.   In comparison to standard alt-az mounts, this mount is expected to be dramatically less expensive, with costs that scale slowly with telescope size.  The telescope needs no standard dome and no massive foundation pier, only shedlike structures covering the two optical elements and possibly some thermal/wind interventions along the horizontal beam path.  Physical access to the mirrors (for cleaning or demounting/recoating) requires no special equipment. 

Due to the large siderostat, WAET systems involve substantial extra mirror area over a conventional telescope.   Most of the extra area is in the form of lower-cost flat siderostat segments rather than figured mirrors.  Second, the primary mirrors are thin (since there is no requirement to resist bending stresses) and have low curvature due to the long focus\cite{stahl_multivariable_2005}, making them likely to be cheaper per unit area than conventional giant telescope mirrors. 

While estimates at this level of detail are necessarily highly uncertain, it is clear that WAET construction costs are very strongly dominated by mirror fabrication costs (with an $A^1$ cost scaling law) and not by components with $A^{>1}$ cost scaling laws.  
(Mount and site engineering are discussed in section \ref{label_mount}.  Cost estimation is discussed in Appendix \ref{label_cost}.)

\subsection{PSF and performance}
WAET allows us to build telescopes whose long dimension is similar to that of large optical interferometers, but with a filled aperture; therefore the resolution and light utilization can be understood via ordinary PSFs.  A few aspects of the WAET PSF differ from other designs and are worth noting.  Other than the central obstruction, the beam path is perfectly clear; there is no secondary-mirror support spider.  The mirrors (particularly the siderostat) are difficult to baffle and may have stray light issues.   (PSFs and stray light are discussed in section \ref{label_optics}.)

\subsection{Design parameter space and variants}  In addition to design flexibility with respect to size and optics, there are many possible variations to the basic steering principle.   Of the three example configurations (dWAET, hWAET, and kWAET) presented in section \ref{label_examples}, hWAET and kWAET have bearing/site/steering layouts as discussed above.   Some alternatives include: 

\begin{itemize}
\item The primary and secondary mirrors could be fixed, and the siderostat could be alt-az mounted.   Compared to the nominal design, this trades a (fairly small) primary-mirror bearing surface for a (quite large-area) siderostat bearing surface.   This might be compatible with circumpolar sky coverage; it might allow cost savings; and it might allow the instrument to fit on a site whose north-south dimension is constrained.   This configuration is suggested for the small telescope (dWAET) discussed in section \ref{section_dWAET}.
\item For a spherical-primary telescope, we could fix the primary mirror to the ground.  In this case, the siderostat still selects $\theta$ while a moving secondary (in the style of HET/SALT/Arecibo) is required to track in $\phi$.   This invokes a tradeoff between sky coverage and (possibly substantial) additional mirror area.  
\item The whole observatory (siderostat, primary, secondary, and possibly cameras) could be mounted on a single large azimuth bearing; this recovers the pointing properties of an alt-az mount, although still only the siderostat is required to roll.   This minimizes mirror area (the siderostat does not need to be ``oversized'' against vignetting); it allows siderostat segment boundaries to align with the primary segment boundaries, simplifying the PSF.  It appears to maximize bearing and site complexity.
\item The WAET geometry is, unusually for a large telescope, compatible with large-aperture refracting elements. For a given prescription, the mass of a refracting element\footnote{Yerkes Observatory's 40'' telescope has a \SI{225}{kg} primary; the \num{48}'' Schmidt corrector at the Samuel Oschin Telescope weighs \SI{36}{kg}.}.   scales with telescope size as $A^{1.5}$ or $D^3$.  In a large circular telescope there is no reasonable way to support such an element.  In WAET, the same prescriptions would require only a rectangular slice of a massive lens, which can be segmented for manufacturability and supported at its bottom edge. 
\end{itemize}

WAET installations are modular and expandable.   Once the foundation, shed, and primary bearing have been built, science operations can begin using subsets of the full mirror inventory.   An operating WAET installation can be upgraded in height (say, from \SI{100 x 2}{\m} to \SI{100 x 4}{\m}) by adding new mirror segments.    All instrument-yard space is accessible during observations.

\section{Sky coverage}\label{label_skycoverage}

The telescope's sky coverage depends on the choice of siderostat width and length.   As before, we discuss this in terms of a polar coordinate system aligned with the siderostat roll axis, with polar angle $\theta$ and azimuthal angle $\phi$ and where $\phi=0$ is the horizon the siderostat faces. 

\begin{figure}\begin{center}
    \includegraphics[width=\textwidth]{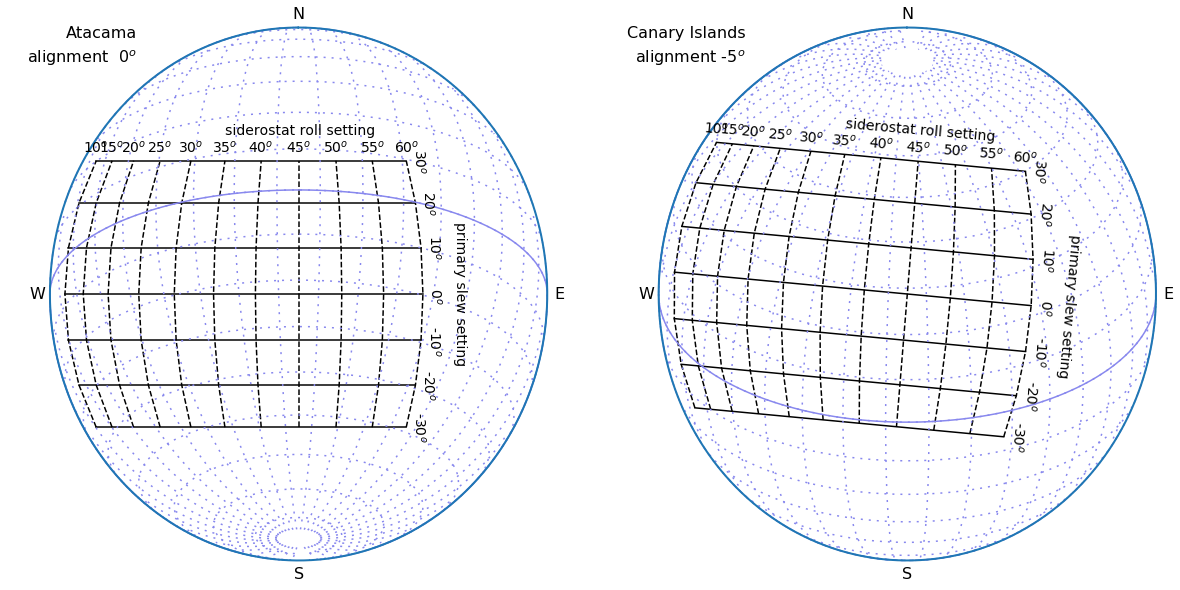}
\end{center}
\caption{Sky coverage and pointing parameters for two illustrative WAET site choices.  The siderostat roll position (where 0$^\circ$=horizon-facing) and the primary slew position (where 0$^\circ$ has the optical axis perpendicular to the siderostat) are shown mapping to orthographic sky coordinates.   Coverage close to the N-S horizons is limited by the $\pm 30^\circ$ slew angle limits, which are likely to be hard mechanical limits.   Coverage of the eastern horizon is shown up to a 60$^\circ$ siderostat roll, which is a vignetting limit that depends on siderostat width.   Left panel: The siderostat long axis is N-S and star tracks are shown for an Atacama site.  Right panel: star tracks are shown for a Canary Islands site, and the siderostat long axis is installed $5^\circ$ E of N to improve field rotation and slew rate behavior.}\label{skycov}
\end{figure}

Except for a target at the forward horizon, the beam encounters the siderostat at an angle $\phi/2$ from the normal, so the siderostat's effective area $A_{\mathrm{eff}}$ is less than its planar area $A$ by $A_{\mathrm{eff}} = A \cos(\phi/2)$.   To avoid vignetting over a sufficient range of rolls, we in general specify an ``overwide'' siderostat.  If the siderostat is wider than the primary by $W_s \sim \sqrt{2} W$, we can point at the zenith ($\phi = 90^\circ$) without vignetting.   A siderostat with $W_s = 2W$ can point from the forward horizon to $\phi=120^\circ$.     Sky coverage in $\theta$ is determined by the slew limits of the primary mirror, and additionally by the projected length of the siderostat.  If the primary mirror is to slew to $\pm 30^\circ$ without vignetting, then the siderostat must be longer than the primary with $L_S =  2/\sqrt{3} L \approx 1.15 L$.   These are the choices whose sky coverage is illustrated in Fig.~\ref{skycov}.

Field rotation is particularly important at WAET due to the asymmetric pupil; interesting features of the field may rotate in and out of alignment with the telescope's high-resolution axis.   This can be read off of Fig.~\ref{skycov}; at any pointing, the telescope's high-resolution axis is parallel to the local slew-coordinate axis, while sky rotation can be seen following the RA/DEC grid.  Even in extreme cases, fields rotate less than $\sim 50^\circ$ over a whole night.   Note that we can choose the telescope's orientation on the ground; if the siderostat is constructed with its (fixed) long axis N-S, the field rotation rate is minimized at zenith.  This may not be optimal; by choosing a different orientation for the fixed axis, we may move the minimum-rotation pointing a bit further east or west, reduce the full rotation range for long observations, and reduce the maximum rate of slew.

Some of the variants mentioned in section \ref{sec_layout} would see more notable rotation effects; these might be beneficial (allowing the high-resolution axis to rotate all the way around a target field which is interesting on both axes) or limiting (reducing the amount of time spent with an optimal view of a particularly interesting axis).   

\section{Mount and site engineering}\label{label_mount}

We argue that WAET systems can be built at extremely low cost.   In conventional telescopes at 8-m and larger scale, the mirrors, mount, and dome/site have roughly equal impacts on the project budget.  WAET's cost savings are driven by the extreme simplification of its mount and dome.

WAET's mirror mechanical supports are much lighter and less complex than an alt-az fork and tube.  The primary mirror segments are mounted on simple rigid steel frames which lend themselves well to mass-production, transport, and installation.  For the baseline hWAET \SI{100 x 2}{\m} aperture, the primary mirror segments are \SI{2 x 2}{\m} squares.  At this scale, complete segment/frame assemblies are light enough for a pallet jack and small enough to fit into standard shipping containers.   The \SI{120 x  2.8}{\m} siderostat is similarly manufactured as 40 identical \SI{2.8 x 3}{\m} segment/cell/bearing/pedestal assemblies, each of which is container- and forklift-compatible.   A preliminary design exercise implemented all mirror supports with \SI{60}{\Hz} stiffness using less than \SI{70}{\tonne} of structural steel.   We note that all mirrors are easily accessible from the back (for, e.g., cooling) and from the front (for cleaning/inspection), removing several complex constraints/interdependencies from the engineering design task.

The cost of site preparation, foundations, and dome are significant cost drivers for large telescopes.   WAET's major elements require only a conventional concrete slab to support them at low load density.   The only site-preparation element whose specifications exceed those of, e.g., warehouse flooring, is primary mirror slew bearing.   The primary mirror support frame assembly needs to slide along a flat surface which is supported by the foundation.  Any irregularities in the bearing surface will be transmitted to platform flexure\footnote{For a conventional telescope, azimuth bearing out-of-flatness applies bending forces to the fork but not to the tube, since it is isolated by the altitude bearings.} (including both decenters and tilts) and require correction by actuators during tracking; in contrast to conventional telescope figure control, these corrections are low-amplitude ($\sim\!$ \SI{10}{\micro\m}) and slow ($\sim\!$\SI{1}{\micro\m\per\hour}), but nonetheless the primary slab flatness and stability requirements need careful consideration.  Given an adequate bearing surface, conventional air bearings\footnote{Roller bearings or oil bearings appear to be overkill for WAET's low load densities.} could be used; alternatively, without engineering a particularly flat surface, the primary mirror could be supported buoyantly, although in this case thermal gradients still cause flexing.   

Although we have portrayed WAET as built a level site\footnote{Flat installations are clearly possible with minimal grading at Llano de Chajnantor, Magdalena Ridge, or at the 100~m scale at Cerro Paranal.}, almost all of the same engineering principles apply if WAET is built on a hillside or slope.  On a sloped site the primary mirror system will see small gravity vector variations during slews (which may drive somewhat higher stiffness specifications); the primary drive and bearing system need to support additional forces.   

In place of a conventional dome, WAET's large mirrors require very simple and conventional shelters---either sheds that roll away during observations, or fixed buildings with retractable walls or roofs.  One site/optics question requiring further study: WAET has a long light path close to and parallel to the ground; ground/air temperature differences and/or turbulent wind on this path would be expected to lead to extremely poor seeing.  On one hand, we have access to particularly-powerful wavefront sensing for this turbulence, described in section \ref{horizontal_ao}.   On the other hand, it will still be desired to minimize the correction amplitudes.  The horizontal beam path is open and naturally well ventilated, but the ground has a large thermal mass.  First, in most cases one wants to elevate the beam path as far as is economical; at minimum by putting the siderostat and primary slabs on berms, or positioning them around a natural depression.  The beampath can of course be elevated by simply building taller frames.  If the dominant turbulence source is convection, the ground's effective thermal mass can be reduced by laying an insulated membrane over the ground under the beam.  If the dominant turbulence source is wind, the appropriate intervention (at more significant, but not prohibitive, cost) is a low roof over the beam path.  This could be accomplished with or without support pillars in the beam path.

\section{Optical performance}\label{label_optics}

The basic layout of WAET is compatible with any conventional optical prescription.  Different choices have different impacts on the size of the primary bearing, the size and existence/nonexistence of the secondary bearing, and the degree of central obscuration.   Here, we show some parameters of a more detailed optical design for a Ritchey-Chr\'etien version of hWAET.

\begin{table}
\begin{center}
\begin{tabular}{l l l r r } 
\hline
\multicolumn{5}{c}{\bf{Telescope overall characteristics}}\\
\hline
\multicolumn{3}{l}{Primary mirror focal ratio} & \multicolumn{2}{r}{ 1.1 } \\
\multicolumn{3}{l}{Telescope focal ratio} & \multicolumn{2}{r}{ 27 } \\
\multicolumn{3}{l}{Unvignetted field of view} & \multicolumn{2}{r}{\SI{30}{\arcmin}} \\
\multicolumn{3}{l}{Scientific field of view} & \multicolumn{2}{r}{\SI{20}{\arcsec}} \\
\multicolumn{3}{l}{Plate scale} & \multicolumn{2}{r}{\SI{13.4}{\milli\m\per\arcsec}} \\
\multicolumn{3}{l}{\SI{600}{nm} resolution element} & \multicolumn{2}{r}{\SI{16}{\micro\m}} \\

\multicolumn{3}{l}{Central obscuration} & \multicolumn{2}{r}{2\%} \\
\hline

\multicolumn{5}{c}{\bf{Mirror specifications}}\\
 & Name & Shape & Curvature (m) & Dimensions (m) \\ 
\hline
M1 &  Siderostat & flat          &  &   \num{120 x 2.8}  \\
M2 & Primary & hyperbolic       &  217 & \num{100 x 2}\\
M3 & Secondary  & hyperbolic &  18 & \num{8 x 0.5}\\
M4 & Fold & flat                     &  & \num{4 x 0.5}\\
M5 & Selector & flat               &   &\num{5 x 0.5}\\
\hline
\end{tabular}
\caption{Optical prescription studied for hWAET.}\label{optical_prescription}
\end{center}
\end{table}

WAET would be the first telescope with sequential segmented mirrors, and the authors are not aware of published cophasing algorithms for such a system.  However, we note that WAET can do autocollimation tests in-situ.  When the siderostat is tilted vertically and the primary is at its center slew position, light emitted from the focal plane returns there after striking M2, M1, and M2 again.  We believe that rapid alignment and cophasing should be possible even with three or more segmented mirrors.  

\subsection{Aperture shape and diffractive PSF} 

Consider WAET's length $L$ and width $W$ as an envelope in which we wish to fit a mirror.   The rectangular aperture is the shape with the best point-source separation ($\theta_R = \lambda/L$) and the largest area $A=L\times W$, and which affords the simplest fabrication; however, it has a notably wide diffraction pattern.    In the same envelope, an elliptical aperture has a lower area $A=0.79 L\times W$ and worse point source separation ($\theta_R = 1.22 \lambda/L$) but more steeply-falling wings.  A diamond-shaped aperture (generalizing from proposed square apertures \cite{lyon_extra-solar_2002, gezari_expo:_2003}) has particularly low wings, but with $A=0.5 L\times W$ and $\theta_R = 1.414 \lambda/L$.  Of course each scientific instrument will have its own aperture stop.   Should we consider making WAET's primary pupil non-rectangular?  If most instruments are expected to mask away the corners of a rectangular aperture, it might be possible to save money by not building out mirror area in those corners to begin with.  On the other hand, these savings might be cancelled out by the loss of uniform mass production.   For the purpose of this paper, our mechanical designs all show a rectangular aperture.  

\begin{figure}
  \begin{minipage}[c]{0.6\textwidth}
    \includegraphics[width=0.8\textwidth]{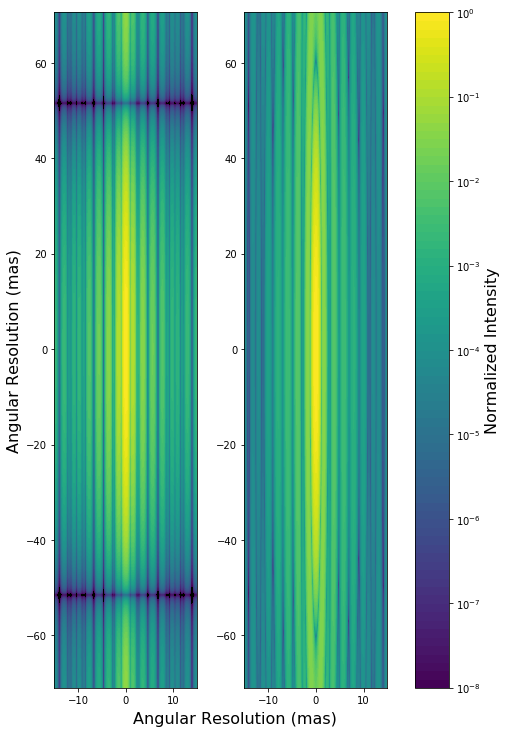}
  \end{minipage}\hfill
  \begin{minipage}[c]{0.4\textwidth}
    \caption{
Monochromatic PSF of \SI{100 x 2}{\m} hWAET at \SI{500}{\nm} showing the extreme resolution asymmetry.  Left: rectangular aperture, including \SI{6}{\mm} segment gaps.  Right: elliptical aperture.
    } \label{psf}
  \end{minipage}
\end{figure}

Successful coronography and planet-finding depends on both the dynamic AO and the static speckle pattern of the telescope.  WAET has several details which simplify the static and quasistatic diffraction pattern.  WAET has no support spider, so the underlying static PSF is very smooth at low orders as seen in Fig.~\ref{psf}.   Although WAET is segmented, since segment motion and mount flexing is almost absent, the segment gap size can be safely pushed to smaller values than on more flexible telescopes\cite{mast_keck_1988}.  Finally, with easy human access to the full installed array and built-in autocollimation optics, new figure-correcting interventions might be possible.  

One factor complicates the PSF: the siderostat and primary mirror each have their own segment-gap and misfiguring PSFs; since the siderostat misfigures are encountered at slowly varying angles, the static speckle pattern in fact changes slowly (but predictably) over the course of an observation (but in a manner more reproducible than what are normally called ``quasistatic'' speckles.)  On one hand this complicates PSF subtraction and deconvolution; on the other hand it may serve as the WAET analogue to angular differential imaging \cite{marois_angular_2006}.

\subsection{Stray light} 

In conventional telescopes, stray light sources include light undergoing diffuse reflections from (a) the mirrors and (b) the telescope structure.  WAET is able to remove 100\% of the structure-related light by removing all mechanical structure from the light path.   However, WAET's M2 and M3 face the horizon; horizon-associated skyglow (or even moonlit ground) can enter the instruments by small-angle diffuse scattering.   A cold stop, extended several times $W$ above the primary and the siderostat, could block this diffuse horizon view. The cold stop can be, for example, (a) a black surface at ambient temperature, (b) flat mirrors reflecting unfocused zenith sky, or (c) low-cost spherical mirrors focused on a colder-than-ambient blackbody cold stop.   The convection-suppressing beampath roof mentioned in section \ref{label_mount} would also serve as a baffle for M2.    Most reasonable shed design will leave M1 exposed to a larger fraction of the sky than a conventional domed telescope's primary, so we expect worse than usual sensitivity to moonlight.   One factor mitigating stray light is that WAET's primary mirrors are unusually accessible for frequent cleaning or recoating.

\subsection{Seeing, adaptive optics and coronagraphy}\label{ao}
A full adaptive-optics design study is beyond the scope of this paper; we will comment on some aspects of WAET that make it differ from other giant telescope AO efforts.

\subsubsection{Horizontal beam AO}\label{horizontal_ao}
WAET has one specific challenge related to ground seeing.   The beam path between the siderostat and primary, traversed at least twice, will encounter substantial turbulence, especially if not enclosed.   If conventional AO techniques (natural and laser guide stars) were required, this might be the single dominant performance limitation of WAET.  However, we have access to both sides of this air volume and can do high-frame-rate, high-fidelity sensing of its contributions to wavefront distortions.  

One implementation of this sensing system is as follows.  Just below the siderostat we place an array R1 of small flat retroreflectors facing the primary.  We place a coherent, narrowband light source at the focal plane.  It is possible to bounce this light through the entire system (M5, M4, M3, M2, R1) so that it returns to the focal plane (R1, M2, M3, M4, M5), traversing the same near-ground path as starlight but twice.  High-quality wavefront sensing or interferometric data from this bright return light is fed into the AO control system.   The system can be duplicated both above and below the siderostat and possibly also along the centerline, either through segment gaps or in dedicated optical access holes, to cover the full 3D beam path.  Multiple wavelengths can correct for chromaticity.   In wind-dominated situations, where phase screens are moving sideways through the system, this AO loop can use predictive control to reduce the effect of servo lag.  

\subsubsection{Conventional AO} For turbulence above the siderostat, WAET's AO capabilities are in principle similar to those of other giant telescopes.  The siderostat itself is somewhat closer to the ground than is typical for modern observatories, and therefore deeper into low turbulence layers.   Three aspects that are unfamiliar from the \SIrange{10}{30}{\m} class are: \begin{enumerate}\item linearly larger plate scales affect the implementation of wavefront sensors, particularly for MCAO. \item Deformable mirrors need to correct larger wavefront phase variance from larger-scale turbulence \item When wind is moving parallel to the narrow axis of the aperture, the AO loop cannot anticipate them with predictive control algorithms; if this is a hard limit, WAET might not progress past the contrast ratios floors associated with frame rates and servo lag.  Multiconjugate AO on widely-separated guide stars can, at least at higher altitudes, sense turbulence outside of the narrow science beam, possibly recovering some predictive capabilities.
\end{enumerate}

\section{Example configurations}\label{label_examples}

\begin{table} 
\begin{center}
\begin{tabular}{ l   r   r   r  }
\hline
\multicolumn{4}{c}{\bf Three example configurations}\\ 
\hline
      & dWAET & hWAET & kWAET \\
\hline
Length (m) & 10.0 & 100 & 300 \\
Width (m) & 0.30 & 2.0 & 5.0 \\
Aperture (m$^2$) & 3.0 & 200 & 1500 \\
$\lambda$/d @1$\mu$m (as) & 21m & 2.1m & 690$\mu$ \\
\hline
\multicolumn{4}{l}{Cost estimates (\$) }\\ 
\quad Primary mirror & 1.50M & 100M & 75M \\
\quad Siderostat mirror & 250k & 17.0M & 127M \\
\quad Supports & 48000 & 3.2M & 24M \\
\quad Foundation/sheds & 15000 & 760k & 5.1M \\
\quad Thermal & 4000 & 10.0M & 18.0M \\
\quad Other costs & 540k & 36M & 65M \\
\hline
Total & 2.3M & 155M & 280M \\
Cost/m$^2$ & 770k & 770k & 188k \\
\hline
\end{tabular}
\end{center}
\caption{Summary of telescope example configurations and scaling-law cost estimates.  dWAET is configured as a low-cost ``trailerable'' telescope for small observatories; hWAET is sized for exoplanet discovery and spectroscopy at a well-studied size scale; kWAET is intended to show how WAET designs scale past \SI{100}{\m}.   There is no obviously-insurmountable barrier to even larger instruments.}
\end{table}

\subsection{hWAET: \SI{100 x 2}{\m}}

\begin{figure}
\begin{center}
\includegraphics[width=\textwidth]{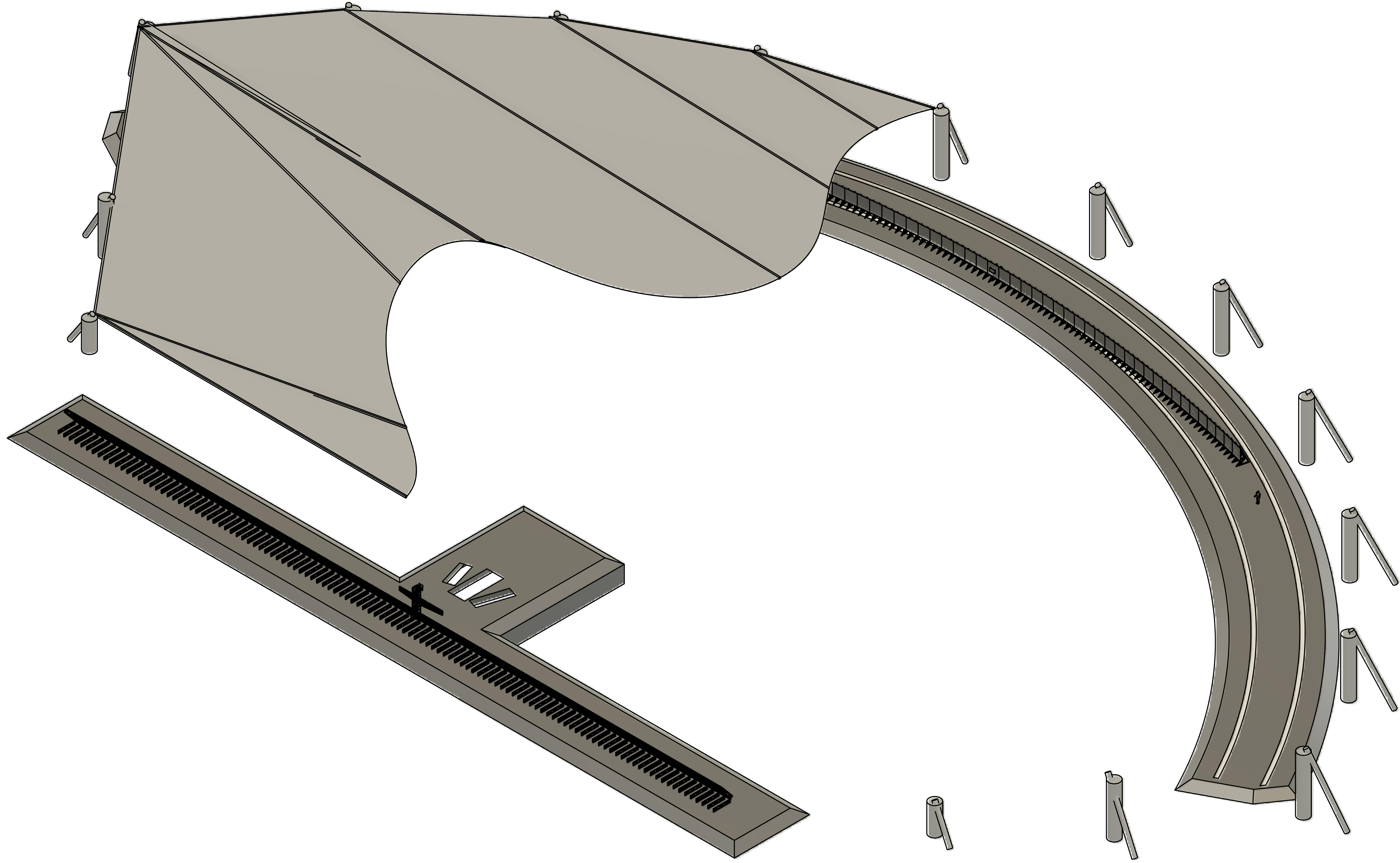}
\caption{Isometric sketch of an enclosed hWAET installation.  At the bottom is the \SI{115 x 3}{\m} siderostat; at the upper right is the \SI{100 x 2}{\m} parabolic primary mirror on its long bearing platform.  Slightly visible here are the tower holding the wide Richtey-Chr\'etien secondary and the instrument selector (center of siderostat); the fold mirror (center of primary); and a \SI{1.8}{\m} person for scale (near right end of primary).   For a detailed view of components see Fig.~\ref{fig2}.   A tension-supported roof (shown cutaway) is stretched over the whole beam path, obscuring the sky within \SI{15}{\deg} of the horizon.}\label{fig_hwaet_mech}
\end{center}
\end{figure}

A \SI{100 x 2}{\m} aperture (hectometer WAET or hWAET)  is the system which, we argue, takes advantage of the WAET layout and realizes key science capabilities, but otherwise has low R\&D risks and does not exceed precedent; in project management terms hWAET has an attractive scope, budget, and timeline.  Figs. \ref{fig1}, \ref{fig2}, and \ref{fig_hwaet_mech} show hWAET in a Richey-Chr\'etien configuration with an f/1.1 primary, f/27 secondary, and instrument rooms below the beam plane.  hWAET's \SI{200}{\m\square} collecting area (equivalent to a \SI{16}{\m} circular aperture) and \SI{2}{\mas} diffraction limit at \SI{1}{\micro\m} (compare to TRAPPIST-1b at \SI{3}{\mas}) are both attractive for exoplanet imaging among other topics.  

\subsection{kWAET: \SI{300 x 5}{\m} spherical} 

\begin{figure}
\begin{center}
\includegraphics[width=\textwidth]{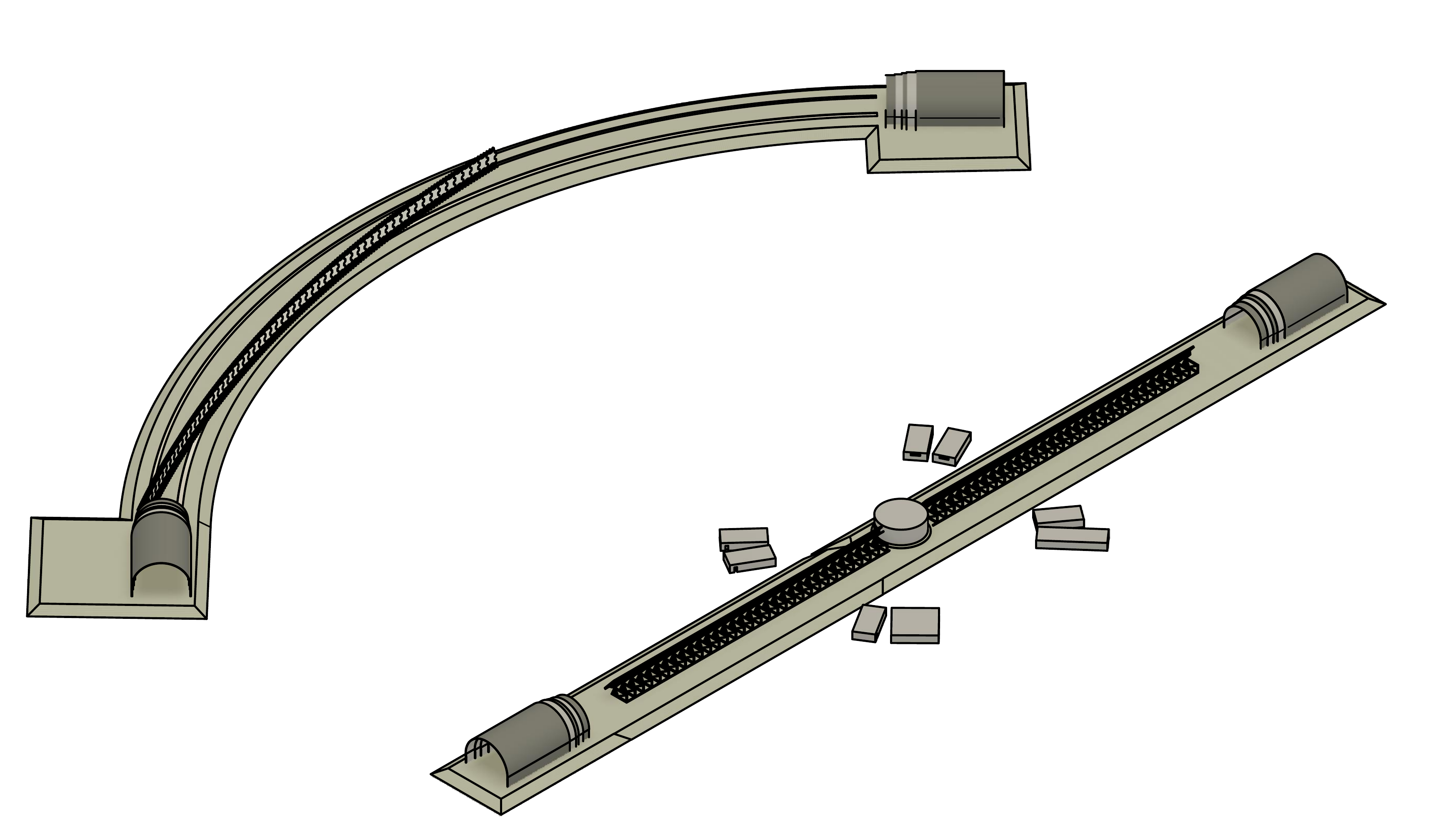}
\caption{Isometric sketch of an open-air kWAET installation.  At the top is the curved bearing slab that supports the slewing motion of the \SI{300 x 5}{\m} spherical primary; the primary is shown at 15$^\circ$ slew to the left.   At the bottom is the foundation slab that supports the tilting \SI{345 x 8}{\m} siderostat.  In the center of the siderostat is a circular shed housing the spherical aberration corrector package, which pivots along with the primary.   Camera buildings (eight shown) may occupy ground space in front of and/or behind the siderostat.  The mirror enclosures are the half-round sheds parked in a nested position at the ends of the slabs. }\label{fig_kraft}
\end{center}
\end{figure}

There is no obviously-insurmountable barrier to a WAET telescope approaching kilometer scale (kWAET).  Consider kWAET to have a \SI{300 x 5}{\m} aperture as in Fig.~\ref{fig_kraft}.  This matches the collecting area of a \SI{50}{\m} class telescope (or 3x TMT) but with a sub-mas diffraction limit.  It is suitable for, among many projects: imaging planets at \SI{1}{\AU} over the entire Kepler field; doing spectroscopy on close-in planets like TRAPPIST-1a; resolving substructure in the first galaxies; resolving surface features on KBOs; etc..   ESO's optical design for OWL supplies us with a cost-saving spherical prescription, which we assume can be followed but without the large central obstruction.  The primary is made of 477 \SI{2.2}{\m} hexagonal segments (probably easier to polish/test than squares) in a \num{3 x 159} grid, mounted on 53 identical nine-mirror subassemblies.  The siderostat segments are 138 \SI{3.4 x 5}{\m} rectangles, installed on 69 identical \SI{6.8 x 5}{\m} siderostat subunits.   The largest corrector-package aspheric mirrors are of order \SI{24 x 0.4}{\m} and necessarily segmented.
  
At the scale of kWAET, project cost estimates are very sensitive to assumptions about economies-of-scale which may or may not be realizable in practice.  However, we can place fairly reasonable upper and lower bounds; the basis for these estimates is detailed in Appendix \ref{label_cost}. For an extreme upper bound, by using an aspheric mirrors and a full tensile roof over beampath, we reach a project cost of \$1.15B.  However, our nominal design has a much lower cost spherical primary.  Using the the spherical mirror production costs estimated for OWL\cite{1998SPIE.3352..778G} and for HET-like ELTs at \SI{30}{\m}\cite{sebring_extremely_1998} or \SI{100}{\m}\cite{ray_issues_2000}, we estimate kWAET could be built for \$210M to \$280M.  

\subsection{dWAET: \SI{10 x 0.3}{\m}}\label{section_dWAET}

\begin{figure}
\begin{center}
\includegraphics[width=\textwidth]{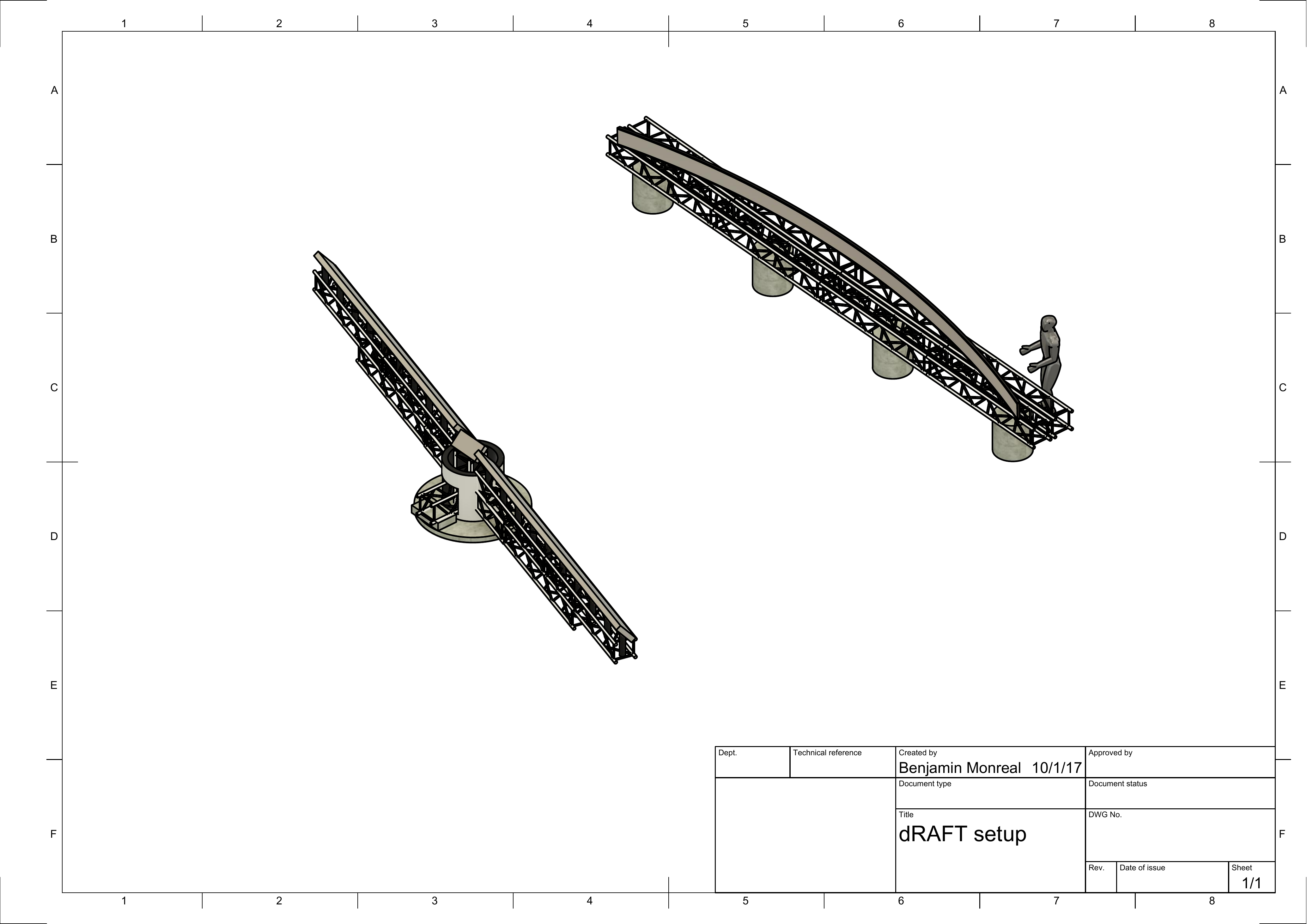}
\caption{Isometric sketch of dWAET.   At upper right is the \SI{10}{\m}$\times$\SI{30}{\cm} Newtonian primary on a stationary \SI{10.6}{\m} truss.  At lower left is a \SI{12}{\m} siderostat mounted on a pivoting truss, shown pivoted by 10$^\circ$. A diagonal flat above the pivot bearing brings the focus to a stationary focal plane. A \SI{1.8}{\m} person is included for scale. Sheds and detailed mirror engineering are omitted.}\label{fig_draft}
\end{center}
\end{figure}

\SI{1}{\m}-class telescopes are now used routinely at small observatories and are associated with stellar astronomy, time domain astronomy, instrument development, and other topics.   By ``flattening'' such an instrument to \SI{10 x 0.3}{\m} (decameter WAET or dWAET) we obtain an instrument with access to \SI{10}{\m}-class resolution on one axis, but potentially with construction/engineering costs more comparable to a conventional \SI{2}{\m}.   Above and beyond conventional small-telescope science, dWAET would add science targets that benefit from its single high-resolution axis; these might include high-redshift galaxy kinematics and minor planet astrometry. A sketch of decameter-WAET (dWAET) is shown in Fig.~\ref{fig_draft}.  It uses a prime-focus Newtonian design for simplicity and ease of alignment.   The \SI{10}{\m} length allows the optics to be mounted on \SI{10}{\m} and \SI{12}{\m} trusses, which fit fully assembled in a standard 40' shipping container.   For ease of site preparation we opt for the fixed-primary, pivoting-siderostat configuration.  One truss, cantilevered from a central pivot bearing, holds the siderostat.  The second, ground-fixed truss holds the primary mirror.  AO is provided by an active flat fold mirror in front of the siderostat; the focal plane is upward-facing inside the siderostat pivot housing.   

\section{Conclusion}

This paper has presented the design of a Wide Aperture Exoplanet Telescope (WAET).  The WAET design is intended to allow future telescopes to scale up to heretofore-infeasible apertures.   We argue that having \emph{one} high-resolution aperture dimension, rather than two, is adequate for a wide variety of science goals, particularly for exoplanet direct imaging.  In addition to general principles of the WAET layout, we show optical prescriptions, mechanical-engineering sketches, and rough cost estimates for instruments with \SI{10}{\m}, \SI{100}{\m}, and \SI{300}{\m} aperture lengths.   

\section{Acknowledgements}

The authors are grateful to Todd Boroson, Corbin Covault, and Ben Mazin for useful discussions, and to Aviv Albeg and Bhavna Nayak for early engineering and design work.

\appendix
\section{Cost estimation}\label{label_cost}

In this paper we have argued for WAET's cost advantages over other giant telescope layouts.  In support of this argument, we attempt to estimate the costs and cost-vs-size scaling laws for WAET components, in some cases based on engineering designs; in some cases on published cost-listing documents from other telescopes and proposals; in some cases from comparable off-the-shelf and commercial prices.  All costs are inflation-adjusted to 2017 USD.  We express scaling laws in terms of the aperture long dimension $L$ and narrow dimension $W$ (both in meters) and focal ratio $N = f/L$.  We restrict the scope to WAET's mechanical components (mounts, cells, bearings, domes) as installed on a flat site.  We do not attempt a detailed accounting of instrumentation (including AO), highly site-dependent grading, outbuildings, management \& overhead, contingency, etc.. but all total-cost figures include a 30\% overhead.   

Despite the roughness of these estimates,  this exercise supports the conclusion that WAET telescope budgets are, unlike those of conventional giant telescopes, almost entirely driven by mirrors.  All other listed engineering work is a small ($<10\%$) correction to the mirror budget.    Therefore, in contrast to conventional telescopes whose cost scales as $A^{\sim 1.35}$, we have shown that WAET costs scale as $A^{\sim 1.0}$ before any economies-of-scale, and with a full manufacturing plan should scale as $A^{< 1.0}$.   If we take the numbers below at face value, we can further state that the ``crossover point'' between WAET cost per area and conventional alt-az cost per area is somewhere around \SI{30}{\m\square}, i.e., that an aspheric WAET telescope of size \SI{0.75 x 40}{\m} would have similar budgets to conventional segmented-mirror telescopes at \SI{6}{\m} diameter; above this size, a WAET mount has a lower cost per area than an alt-az mount.  Therefore, WAET is a cost-effective way of building a light bucket, even before the advantages of its high-resolution axis are exploited.    WAET is of course far, far cheaper than a circular filled aperture of with the same diffraction limit. 
  
\paragraph{Mirrors}  Interpolating from publicly-available data, we price thin aspheric mirrors at \$450k/m$^2$ and flat mirrors at \$50k/m$^2$, we can estimate the total mirror cost per unit WAET aperture to be \$580k/m$^2$ for an aspheric primary.    For spherical primaries, OWL estimates\cite{1998SPIE.3352..778G} allow us to calculate spherical mirror costs in mass production including the effect of blank thickness leading to a cost per unit WAET aperture of \$50--\$90k/m$^2$ depending on primary and siderostat thicknesses.   

\paragraph{Mounts \& bearing} From preliminary engineering designs, we have estimated the mirror support frame costs scale roughly linearly with aperture.  (In more detail, increases in $L$ drive slower-than-linear cost increases due to economies of scale; taller structures probably make cost increase worse-than-linearly with $W$.)  A design exercise at hWAET scale required 70~t of structural steel, which if fabricated at ~\$10/kg suggests a cost per aperture of \$3500/m$^2$.  High-precision air bearings (linear for the slew, rotary for the siderostat tilt), priced at ~\$5000 per tonne of load capacity, add \$2000/m$^2$ per aperture; these may be overspecified for the job so we treat this as an upper bound.   Each mirror unit requires a support cell and figure-control actuators; OWL estimated\cite{1998SPIE.3352..778G} \$25k/m$^2$ for such cells; WAET designs need only thermal and bearing-shape corrections and with a known gravity vector, ought to be far less expensive per unit mirror; but the mirror area is $2.4\times$ the aperture area; a very rough estimate is \$10k/m$^2$ per unit aperture.

\paragraph{Foundations and sheds} The siderostat and primary mirror rest on concrete slabs with load requirements comparable to, e.g., commercial warehouse flooring.  For the nominal steering mechanism, with a slewing primary, the total slab area is roughly $0.75 (L/\mathrm{m})^{1.6}$~m$^2$ and is very weakly dependent on the focal ratio $N$ or width $W$.  For an alt-az siderostat, the slab area is roughly $L^2/2$.  We estimate \$300/m$^2$ (a factor of six more than conventional warehouse flooring) would cover the remote site and, in the case of an air bearing, an unspecified method for obtaining the desired flatness.  Instead of an enclosure dome, garage-like shelters can roll over the mirrors; from prefab steel building catalog prices we estimate the shelter cost as \$2000 $ (L/\mathrm{m}) \times  (W/\mathrm{m})^{0.5}$.  Such sheds may not be necessary if the thermal-intervention includes a full roof.  

\paragraph{Beampath thermal interventions} We have identify three possible interventions which might be necessary to suppress ground- and/or wind-related seeing along the long beampath between M1 and M2.  \begin{itemize} \item To mitigate convection but not wind-related turbulence, we have priced a membrane or plenum, laid on a layer of insulation on the ground and designed to equilibrate quickly with the air; based on commercial roofing costs we estimate \$$40 \times (N (L/\mathrm{m}))^{2}$.  \item A flat steel roof over the beampath, with unsupported spans of $<$\SI{20}{\m} and vertical columns in the beampath as needed.   Based on commercial steel building costs we estimate \$200$ \times (N (L/\mathrm{m}))^{2}$.  \item A roof over the beampath without in-beam pillars requires long unsupported spans, perhaps tensile, and is more difficult to price with any confidence.  Extrapolating from some stadium and hangar projects, we estimate such a roof at \$1k$ \times (N (L/\mathrm{m}))^{2}$ although here the exponent may be greater than 2.  Note that long focal ratios require very large roofs. \end{itemize}
 
\paragraph{Other components} 
We do not attempt detailed cost breakdowns for: support buildings, software, electronics, metrology, instruments, adaptive optics, the ground-layer beampath wavefront measurement system, engineering design services, mirror temperature control and cleaning, or management and operations, or contingency.   Based on \cite{sebring_extremely_1998} and \cite{ray_issues_2000} we quote a 30\% overhead for all such costs.

\begingroup\tabcolsep=2pt
\begin{table}
\begin{center}
\begin{tabular}{ l  r  l  r  l } 
\hline
\multicolumn{5}{c}{\bf{Cost scaling estimates}}\\
\hline
\multicolumn{2}{c}{Component} & \multicolumn{1}{l}{ Version} & \multicolumn{2}{l}{Scaling estimate}\\
\hline
Steel framework & &          & \$3k & L$\times$W\\
Air bearings & & & \$2k & L$\times$W\\
Mirror cells & &   & \$10k & L$\times$W\\
Enclosures & &  & \$2k & L$\times$W$^{0.5}$\\
Siderostat mirror & &       &  \$58k  & L$\times$W\\
\addlinespace

 \multirow{2}{1.1in}{Primary mirror}  &  \multirow{2}{0.1in}{\raggedleft $\bigg\{$}       &  Aspheric & \$450k  & L$\times$W\\
 & & OWL-like & \$50k &  L$\times$W\\
\addlinespace
\multirow{2}{1.1in}{Slab and bearing} & \multirow{2}{0.1in}{\raggedleft $\bigg\{$}  & Primary-slew     & \$250 & L$^{1.6}$\\
 &&  Siderostat-pivot & \$150 & L$^{2}$ \\
\addlinespace
 \multirow{3}{1.1in}{Thermal control}  & \multirow{3}{0.1in}{\raggedleft  $\Bigg\{$} & Ground cover   & \$40 & L$^2$\\
       &                             & Roof (flat)      & \$200 & L$^2$  \\
        &                            & Roof (tension)      & \$1k & L$^2$  \\
\hline
\end{tabular}
\caption{Summary of estimated cost scaling laws.  Aperture length $L$ and width $W$ in meters, costs in 2017 US dollars.}
\end{center}
\end{table}
\endgroup

\begin{figure}
\includegraphics[width=\textwidth]{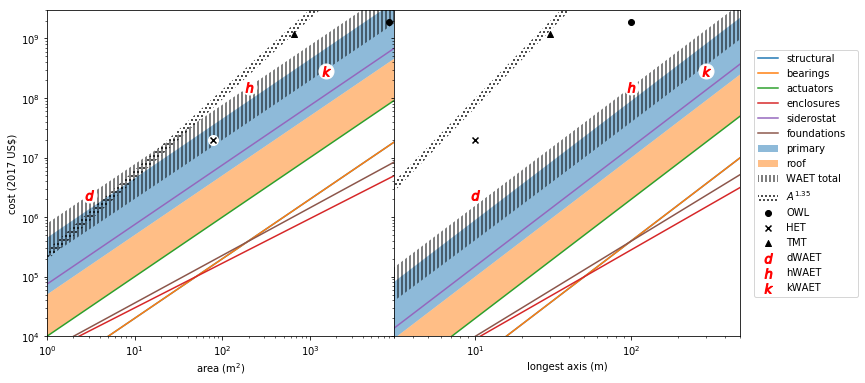}
\caption{Cost estimates and scaling as a function of area (left) or longest-dimension (right), assuming a 50:1 aspect ratio.  The uncertainty bands for the primary mirror shows the range between aspheric and spherical mirrors; uncertainty bands for the roof span the flat and tension options.  For reference, we show cost ballparks for HET, TMT, and OWL as well as projections for dWAET and hWAET (aspheric) and kWAET (spherical).  (Note that kWAET is slightly thinner, and dWAET slightly fatter, than the 50:1 aspect ratio otherwise plotted.)   The dotted band shows an $A^{1.35}$ scaling law, spanning Keck-like or TMT-like reference costs.}\label{fig_cost}
\end{figure}


\bibliographystyle{plain}   

\bibliography{flatscope}   

\end{spacing}

\end{document}